\documentstyle[preprint,aps,epsf]{revtex}
\begin{document}
\baselineskip=15pt 

{\hfill YITP-99-72\ \ }

\begin{center}
{\large {\bf  Current distribution in Hall bars and\\
 breakdown of the quantum Hall effect}}
\bigskip

{ K. Shizuya}
\bigskip

{\sl Yukawa Institute for Theoretical Physics\\
 Kyoto University,~Kyoto 606-8502,~Japan }
\end{center}

\noindent
{\bf Abstract} 
\par
\baselineskip=12pt 
{\small A numerical study is made of current distribution in small
Hall bars with disorder.
It is observed, in particular, that in the Hall-plateau
regime the Hall current tends to concentrate near the sample edges
while it diminishes on average in the sample interior as a consequence
of localization.
Also reported is another numerical experiment on a related, but rather
independent topic, the breakdown of the quantum Hall effect.
It is pointed out that the competition of the Hall field with disorder
in the sample interior, an {\it intra}\,subband process, can account
for both the magnitude and 
magnetic-field dependence $\propto B^{3/2}$ of the critical breakdown
fields observed experimentally.}\\


\baselineskip=13.6pt 

\noindent
{\bf 1. Introduction}
\smallskip

The current distribution in a Hall sample is a subject of interest
pertaining to the foundation of the quantum Hall effect (QHE).
In spite of various studies, both experimental [1] and theoretical
[2-6], it is still a focus of attention and controversy whether the Hall
current flows in the sample interior or along the sample edges. 

In this paper we first report on a numerical study of current
distribution in small Hall samples with disorder and verify
some theoretical observations [7].
In particular, we demonstrate an important physics of redistribution
of the Hall current via disorder. 

A by-product of the numerical analysis is a hint as to a rather
independent subject, the breakdown of the quantum Hall effect [8-13].
We report on a numerical analysis [14] to verify that 
the competition of the Hall field with disorder in the sample 
interior, an {\it intra}\,subband process, can account for 
both the magnitude and magnetic-field dependence $\propto B^{3/2}$ of 
the critical breakdown fields $E^{\rm cr}$ observed in a series of 
experiments by Kawaji {\it et al.} [11-13].\\

\noindent
{\bf 2. Method of calculation}
\smallskip

Consider a Hall bar bent into a loop of circumference
$L_{x}$ and width $L_{y}$, with a uniform magnetic field $B$ normal to
the plane. We distribute short-range impurities over it through 
an impurity potential
$U(x,y) = \sum_{i}\lambda_{i}\, \delta (x - x_{i}) \delta (y - y_{i})$,
and detect the current $j_{x}$ flowing in response to
a uniform Hall field $E_{y}$.
We take the $y=0$ edge to be a sharp edge where the wave function is 
bound to vanish and leave the other edge $y=L_{y}$ as a gentle edge.

For numerical calculations 
it is advantageous to work with the basis $\{ |n,y_{0}\rangle \}$ of 
the eigenstates of the $U=E_{y}=0$ case, labeled
by $n = 0,1,2,\cdots$, and $y_{0}=\ell^{2}\,p_{x}$, with 
${\ell}\equiv 1/\sqrt{eB}$.
Diagonalizing the Hamiltonian with respect to Landau labels $n$ 
by a unitary transformation $W$, one obtains
the Hamiltonian governing each impurity-broadened Landau subband:
\begin{eqnarray}
&&H^{W}_{n}(y_{0},y_{0}')
=\! h(y_{0})\,\delta_{y_{0} y_{0}'}  + U_{nn}(y_{0},y_{0}') +
\cdots, \nonumber\\
&&h(y_{0}) =\omega \{\nu_{n}(y_{0})\!+\!{\textstyle {1\over{2}}} \} 
+ eE_{y} \{y_{0}\!-\!\ell^{2} {\nu_{n}}'(y_{0})\} ,
\label{HWnn}
\end{eqnarray}
where the confining potentials $\omega \nu_{n}(y_{0})$ relevant 
around $y_{0}\sim 0$ derive from the sharp edge.

Numerically diagonalizing $H^{W}_{n}$, one obtains eigenstates
$|\alpha\rangle$ forming the $n$th subband and can calculate
the associated current density $j_{x}^{(\alpha)}(x,y)$.
We shall distinguish between $j_{x}^{(\alpha)}(x,y)$
present even in equilibrium and its response to a Hall field,
the Hall-current component $j_{\rm Hall}^{(\alpha)}(x,y) = 
j_{x}^{(\alpha)}(x,y|E_{y}\!+\!\delta E_{y}) 
- j_{x}^{(\alpha)}(x,y|E_{y})$, 
calculated with $\delta E_{y}= E_{y}/100$.

There are three competing effects to be considered. 
(i) Impurities capture electrons and make them localized. 
In contrast, (ii) a Hall field makes them drift with
velocity $v_{x}\sim E_{y}/B$.
A simple estimate of energy cost
reveals how these two effects compete:
An electron state localized around an isolated
impurity of strength $\lambda$ acquires an energy shift 
$\triangle \epsilon \approx \lambda/(2\pi \ell^{2})\equiv s\,\omega$,
where we have introduced a dimensionless strength $s$. 
When a Hall field is turned on, it will perceive over its spatial
extent of $O(\ell)$ an energy variation of 
magnitude $\sim e \ell\,E_{y}$.
Accordingly, if the field becomes so strong that the field-to-disorder 
ratio
\begin{equation}
R\equiv  e \ell\, |E_{y}|/(|s|\, \omega) \gtrsim 1 ,
\label{svsE}
\end{equation}  
the electron state would be delocalized. 
The numerical experiment given below reveals that 
$R^{\rm \, cr} \sim O(0.1)$ is the critical value for delocalization.
Finally, (iii) the sharp edge $y=0$ drives
electrons along it by an ``effective field'' as strong as 
$e\ell E_{y}^{\rm eff} \sim \omega \ell\,{\nu_{n}}' \sim \omega$. \\

\bigskip

\noindent
{\bf 3. Hall-current distribution}
\smallskip

We have examined current distributions for a number of samples.
Here we mainly quote the case of a sample of length 
$\sqrt{2\pi}\ell \times 28 \approx 70 \ell$ and width
$\sqrt{2\pi}\ell \times 8 \approx 20 \ell$,
supporting $28\times 8 = 224$ electron states 
in the $n=0$ subband.  180 impurities of varying strength 
$|s_{i}|\le s_{\rm max}\sim 0.1$ are randomly distributed on it.
We choose $E_{y}<0$ so that the $y=0$ edge is the ``upper''
edge, and in most cases take it very weak, 
$R=e\ell |E_{y}|/(s_{\rm max}\, \omega) \sim 1/10^{4}$.

We first summarize the observations drawn from our numerical 
analysis [7]: \hfil\break
(i) The equilibrium current $j_{x}$ and 
the Hall current $j_{\rm Hall}$ are substantially different in
distribution.
(ii) In the Hall-plateau regime the edge  
states (of the uppermost subband) are vacant and scarcely
contribute to the current distribution. This indicates that the edge
states are in no sense the principal carriers of the Hall current. 
(iii) In the plateau regime the Hall current $j_{\rm Hall}$
tends to diminish on average in the sample bulk and concentrate 
near the sample edges.
(iv) The sharp edge and disorder
combine to efficiently delocalize electrons near the edge.
(v) The Hall field competes with disorder to delocalize
electrons in the sample bulk.

Evidence for (i) and (ii) is seen from Fig.~1(a), where the total 
currents $J_{x}$ and $J_{\rm Hall}$ carried by the $n=0$ subband are
plotted as a function of vacancies $N_{\rm v}$. 
There the edge states, carrying a large amount of current per state,
are readily identified through a rapid decrease of $J_{x}$ for
$0\le N_{\rm v} \lesssim 15$.
They, however, have little effect on $J_{\rm Hall}$.

Note the contrast between the upper and lower plateaus in Fig.~1(a). 
This suggests observation (iv).

See next the ($x$-averaged) Hall-current distribution across
the sample width in Fig.~1(b), which clearly demonstrates (iii).  
The density distribution in Fig.~1(c) also shows that in the plateau
regime more electrons survive near the sample edge than in the
interior. This gives evidence for the expulsion of
the Hall current out of the disordered sample bulk, expected
theoretically [7]. 
The decrease of the current in the sample bulk is correlated with
the dominance of localized states there.  This is confirmed by
increasing a Hall field or $R$ gradually: 
Drastic changes arise around $R\sim 0.1$; there the Hall plateaus
disappear and a sizable amount of Hall current is seen to flow 
in the bulk. This leads to observation (v).

Figure~2 shows another demonstration of current redistribution 
via disorder.  There it is clearly seen that electron states residing 
on the edges of the disordered region (``bulk edges'') support 
a considerable amount of Hall current per state and that a small number 
of states in the inner bulk carry an even larger amount.

Observation (v) suggests a possible mechanism for the
breakdown of the QHE. We study the competition of the Hall field with
disorder in detail in the next section.\\

\bigskip  

\noindent
{\bf 4. Field-induced breakdown of the QHE}
\medskip

To simulate electrons in the bulk of 
a realistic sample we have considered electron states residing over a
disordered domain of size $\approx 70 \ell \times 30 \ell$, where
360 random impurities with $|s_{i}| \le s_{\rm max}$ are 
distributed [14].
The main observation drawn from our numerical experiment is that,
in the regime of dense impurities $N_{\rm imp}/N_{\rm state}>1$ (which
presumably applies to realistic samples), the number of localized
states, $N^{\rm loc}$, decreases exponentially with $|E_{y}|$ and
obeys an approximate scaling law written in terms of 
\begin{equation}
R \equiv e \ell\, |E_{y}|/(s_{\rm max}\,\omega) 
\propto |E_{y}|/(s_{\rm max}\,B^{3/2}). 
\end{equation}
See the $N^{\rm loc} - R$ plots for the uppermost subband in Fig.~3(b);
they stay approximately the same for a wide variation in magnetic field
$B= B_{0}/\kappa$ with $\kappa=1\sim 4$.
See also the $J_{x}^{\rm Hall} - N_{\rm v}$ plots in Fig.~3(a) for various
choices of $R$. Actually the data refer to the $n=0$ subband, 
but the way the plateaus shrink with increasing $R$ turns out virtually 
the same for higher $n=1,2,3$ subbands as well.
This shows that the field-to-disorder ratio $R$ is a good measure to
express field-induced delocalization of electron states.

The underlying physical picture revealed by this scaling behavior of
$N^{\rm loc}$ is that the electron states remaining localized 
in the near-breakdown regime are always governed by the same set of
impurities of large strength $|s_{i}| \sim |s|_{\rm max}$;
that is, one always encounters the same set of localized states 
prior to breakdown. We have confirmed this by observing that 
suppressing weaker impurities (e.g., $|s_{i}| < 0.5\, s_{\rm max}$)
entails no essential change in the behavior of $N^{\rm loc}$ 
for large $R$.

This picture offers a simple explanation for the observed 
$E^{\rm cr}_{y} \propto B^{3/2}$ law of the critical breakdown field.
(The fact that this scaling law is observed for samples differing in
carrier density and mobility~[11-13] would be attributed to relatively
small variations in disorder-strength $|s|_{\rm max} \sim O(0.1)$ for
a variety of samples.)   
It is also consistent with the observation of Cage {\it et al.}~[9]
that the breakdown of the dissipationless current is spatially
inhomogeneous.
Finally, it is a natural consequence of the present 
{\it intra}\,subband process that the magnitude of $E^{\rm cr}_{y}$
falls within the observed range of a few hundred V/cm, one order of
magnitude smaller than what intersubband processes, such as Zener
tunneling~[10], typically predict. \\

%\bigskip

\noindent
{\bf 5. Concluding remark}
\medskip

As for current redistribution via disorder, discussed in Sec.~3, 
it is essential to separate 
the (slow) Hall-current component $j_{\rm Hall}$ from  
the (fast) equilibrium current $j_{x}$. 
A way to achieve such separation in experiment is to add a small
alternating component to an injected direct (or slowly-alternating)
current. Information on the Hall-current distribution 
would then be obtained by detecting the alternating component in 
the Hall-potential distribution responding to it. \\

\newpage

\noindent
{\bf Acknowledgments}
\medskip

This work is supported in part by a Grant-in-Aid for 
Scientific Research from the Ministry of Education of Japan, 
Science and Culture (No. 10640265).  \\

{\bf References}
\medskip

\noindent
[1] P. F. Fontein {\it et al.}, Phys. Rev. B 43 (1991) 12 090, and
earlier references cited therein.

\noindent
[2] A. H. MacDonald, T. M. Rice, and W. F. Brinkman,
 Phys. Rev. B 28 (1983) 3648.

\noindent
[3] O. Heinonen and P. L. Taylor, Phys. Rev. B 32 (1985) 633.

\noindent
[4] T. Otsuki and Y. Ono, J. Phys. Soc. Jpn. 58 (1989) 2482.

\noindent
[5] D. J. Thouless, Phys. Rev. Lett. 71 (1993) 1879.

\noindent
[6] T. Ando, Phys. Rev. B 49 (1994) 4679.

\noindent
[7] K. Shizuya, Phys. Rev. B 59 (1999) 2142; 
Phys. Rev. Lett. 73 (1994) 2907.

\noindent
[8] G. Ebert, K. von Klitzing, K. Ploog, and G. Weimann,
J. Phys. C 16 (1983) 5441.

\noindent
[9] M. E. Cage {\it et al.}, Phys. Rev. Lett. 51 (1983) 1374.

\noindent
[10] H. L. Stormer {\it et al.}, 
Proc. 17th Int. Conf. Physics of Semiconductors, San
Francisco, 1984, ed. J. D. Chadi and W. A. Harrison (Springer Verlag, 
New York,1985) p.~267.

\noindent
[11] S. Kawaji {\it et al.}, J. Phys. Soc. Jpn. 63 (1994) 2303.

\noindent
[12] T. Okuno {\it et al.}, {\it ibid.} 64 (1995) 1881.

\noindent
[13] T. Shimada, T. Okamoto, and S. Kawaji, 
Physica  B 249-251 (1998) 107.

\noindent
[14] K. Shizuya, Phys. Rev. B 60 (1999) 8218.

%%%%%%%%%%%%%%%%%%%%% Figure captions %%%%%%%%%%%%%%%%%%%%%%%
\begin{figure}
\centerline{\epsfbox{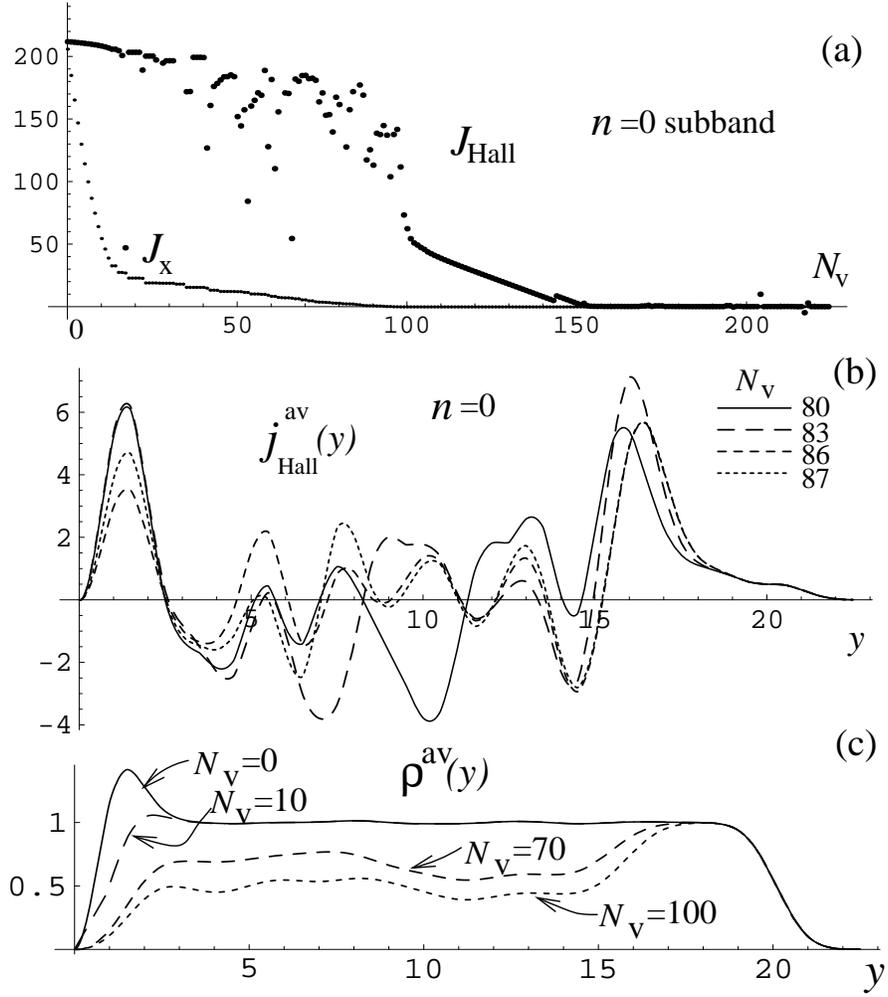}}
\vspace{0.5in}
\caption{ (a) Total currents $J_{\rm Hall}$ and $J_{x}$ 
(on different scales) vs vacancies $N_{\rm v}$.
(b) Hall-current distribution $j_{\rm Hall}^{\rm av}(y)$
[in units of $-(e^{2}/2\pi \hbar)\,\delta E_{y}$] in the plateau
regime. (c) Density distribution.}
\end{figure}
\bigskip

\newpage

\begin{figure}
\centerline{\epsfbox{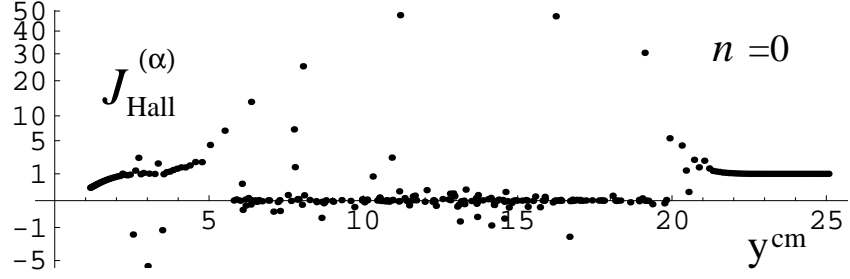}}
\vspace{0.5in}
\caption{Net amount of Hall current per state [on a square-root scale]
as a function of center-of-mass position $y^{\rm cm}$ for each
state. Impurities lie over $5 \ell \le y \le 20 \ell$.}
\end{figure}

\vspace{2cm}

\begin{figure}
\centerline{\epsfbox{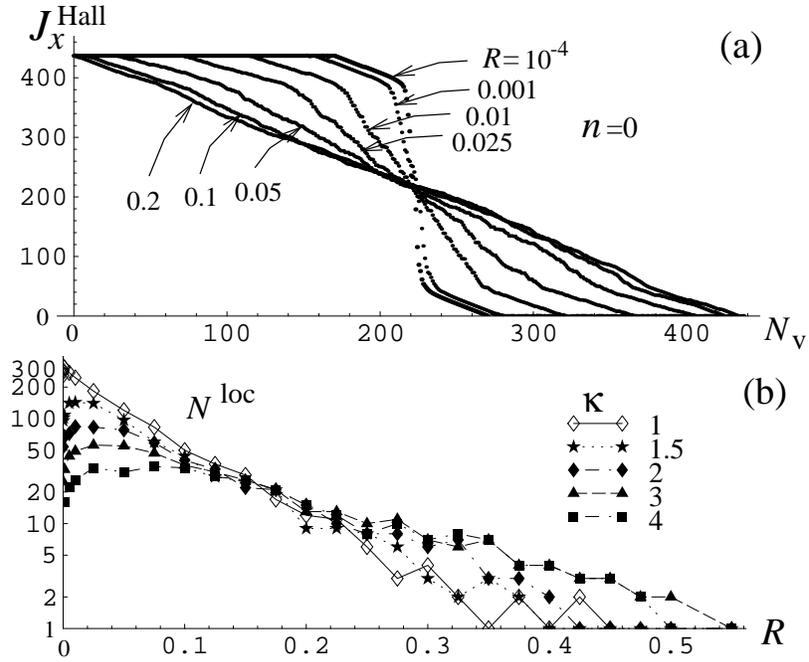}}
\vspace{0.5in}
\caption{ (a) $J_{x}^{\rm Hall}$ vs $N_{\rm v}$ for 
$R= 0.0001 \sim 0.2\,.$\\
(b) $\log N^{\rm loc}$ vs $R$ for $B=B_{0}/\kappa$ with 
$\kappa= 1\sim 4$. }
\end{figure}

\end{document}